\documentclass[aps, prd, 10pt, twocolumn, superscriptaddress,noshowpacs, preprintnumbers, longbibliography, bibnotes,hyperref,floatfix,nofootinbib]{revtex4-1}

\usepackage[colorlinks=true,breaklinks=true]{hyperref}
\hypersetup{allcolors=[rgb]{0.0 0.0 1.0},linkcolor=[rgb]{0.75 0.05 0.05}}
\usepackage[dvipsnames]{xcolor}
\usepackage{mathtools}
\usepackage{amsmath}
\usepackage{amsfonts}
\usepackage{amssymb}
\usepackage{physics}
\usepackage{bbold}
\usepackage{bm}
\usepackage{hyperref}
\usepackage{orcidlink}
\usepackage{soul}

\newcommand{\bD}{{\bf D}}

\newcommand{\be}{{\bf e}}

\newcommand{\bP}{{\bf P}}

\newcommand{\bJ}{{\bf J}}

\newcommand{\bp}{{\bf p}}

\newcommand{\sH}{{\sf H}}
\newcommand{\sP}{{\sf P}}
\newcommand{\sA}{{\sf A}}
\newcommand{\sB}{{\sf B}}

\newcommand{\exclude}[1]{{}}
\long\def\exclude#1{}

\begin{document}

\preprint{MPP-2023-284, SLAC-PUB-17760}

\title{Collisions and collective flavor conversion: Integrating out the fast dynamics}

\author{Damiano F.\ G.\ Fiorillo \orcidlink{0000-0003-4927-9850}} 
\affiliation{Niels Bohr International Academy, Niels Bohr Institute, University of Copenhagen, 2100 Copenhagen, Denmark}

\author{Ian Padilla-Gay
\orcidlink{0000-0003-2472-3863}} 
\affiliation{SLAC National Accelerator Laboratory, 2575 Sand Hill Road, Menlo Park, CA, 94025}

\author{Georg G.\ Raffelt
\orcidlink{0000-0002-0199-9560}}
\affiliation{Max-Planck-Institut f\"ur Physik (Werner-Heisenberg-Institut), Boltzmannstr.~8, 85748 Garching, Germany}

\begin{abstract}

In dense astrophysical environments, notably core-collapse supernovae and neutron star mergers, neutrino-neutrino forward scattering can spawn flavor conversion on very short scales. Scattering with the background medium can impact collective flavor conversion in various ways, either damping oscillations or possibly setting off novel collisional flavor instabilities (CFIs). A key feature in this process is the slowness of collisions compared to the much faster dynamics of neutrino-neutrino refraction. Assuming spatial homogeneity, we leverage this hierarchy of scales to simplify the description accounting only for the slow dynamics driven by collisions. We illustrate our new approach both in the case of CFIs and in the case of fast instabilities damped by collisions. In both cases, our strategy provides new equations, the {\it slow-dynamics equations}, that simplify the description of flavor conversion and allow us to qualitatively understand the final state of the system after the instability, either collisional or fast, has saturated.
\end{abstract}

\maketitle

\section{Introduction}\label{sec:introduction}

Collective flavor oscillations in neutrino-dense environments are driven by neutrino-neutrino refraction that also possesses off-diagonal flavor components~\cite{Pantaleone:1992eq}. The magnitude of the induced refractive energy shift is set by the neutrino number density $n_\nu$, and can roughly be measured by the scale $\mu=\sqrt{2}G_{\rm F} n_\nu$, with $G_{\rm F}$ the Fermi constant; everywhere in this work, we use natural units such that $c=\hbar=k_{\rm B}=1$. For typical conditions of a supernova (SN) and the remnant of a compact object binary, the neutrino-neutrino interaction strength can be as large as $\mu = 10^5 \ \mathrm{km^{-1}}$ in the vicinity of the neutrino decoupling regions and can drive neutrino fast flavor conversions (FFC) that operate on timescales of a few nanoseconds. The timescale $\mu^{-1}$ is much smaller than the one of neutrino slow oscillations $\sqrt{\mu \omega}^{-1}$, where $\omega = \Delta m^2/2E_\nu$; for a typical neutrino energy of 15~MeV and the largest mass-squared difference one finds $\omega\simeq 0.4~\mathrm{km^{-1}}$. Since the potentials span over a wide range, collective flavor oscillation can also span over a wide range of time and length scales, for recent reviews see Refs.~\cite{Duan:2010bg, Mirizzi:2015eza, Tamborra:2020cul, Capozzi:2022slf, Richers:2022zug}.

Recently, there has been a surge of interest in the interplay between the coherent forward scattering $(\propto G_{\rm F})$ and the incoherent scattering of neutrinos off matter $(\propto G_{\rm F}^2)$. Roughly, the ratio of coherent to incoherent rates is given by $\Gamma/\mu \simeq G_{\rm F} E_\nu^2 \sim 10^{-9}$ for a 15~MeV average neutrino energy; the regime where $\Gamma \ll \mu$ is referred to as the weak-damping regime. Otherwise, if $\Gamma \sim \mu$ oscillations are generally damped, similarly to the quantum Zeno effect~\cite{Harris:1980zi}, but such conditions never seem to occur in realistic astrophysical or cosmological environments. 

Conventional wisdom suggests that incoherent collisions should damp the off-diagonal coherence required to exhibit collective motion, and therefore destroy collective oscillations. However, as realized recently, this is not always the case if the collisional rates depend on energy or differ for neutrinos and antineutrinos~\cite{Johns:2021qby,Xiong:2022zqz, Fiorillo:2024, Liu:2023pjw, Lin:2022dek, Johns:2022yqy, Padilla-Gay:2022wck}, which might trigger novel collisional flavor instabilities (CFIs). Even if CFIs have a subdominant role, collisions might still play a role in damping the fast oscillations which likely occur in the inner regions of SNe.

Investigating the impact of collisions on the final state of flavor conversions requires an investigation of the nonlinear evolution of the latter. To tackle this issue, hierarchies among different scales are one key to simplifying the problem. As mentioned before, one such hierarchy is $\Gamma\ll\mu$.
Therefore,  whether collisions drive instabilities or simply damp the collective fast oscillations driven by forward scattering, their dynamics is slow, and the separation of these scales can be used to simplify the problem. 

A similar separation is key to understanding the traditional MSW effect~\cite{Wolfenstein:1977ue,Wolfenstein:1979ni,Mikheyev:1985zog} where fast in-matter oscillations separate from the slow change of density along the beam. Likewise, in a dense medium, a large matter effect is often lumped into a small effective in-medium mixing angle and otherwise ignored because the fast oscillations can be averaged out. On the other hand, for oscillations driven by neutrino-neutrino refraction, such separation-of-scale techniques have not been explored. This is a more challenging scenario because even the rapid oscillations are nontrivial, being driven by nonlinear dynamics.

Here we address precisely this problem. We show that due to the hierarchy $\Gamma\ll \mu$, the dynamics of oscillations impacted by collisions can be separated into a \textit{fast} dynamics, on timescales $\mu^{-1}$, and a \textit{slow} dynamics\footnote{It should not be confused with the slow flavor conversion phenomenon which is characterized by the $\sqrt{\mu\omega}^{-1}$ timescale.}, on timescales $\Gamma^{-1}$. In practice, the separation is performed by averaging over the fast dynamics, and identifying a set of physically relevant quantities that only change slowly. This allows us to obtain a new set of equations, henceforth referred to as \textit{slow-dynamics equations} (SDEs), which do not contain terms of order $\mu$. Therefore, solving the SDEs requires to follow the dynamics only over timescales of order $\Gamma^{-1}$, without keeping track of the fast motion which is not relevant to the final flavor outcome of the evolution. Admittedly, these new equations are still nonlinear and coupled; yet, they can be solved much faster than the original equations of motion (EOMs), while offering significant advantages in terms of intuitive understanding.

We apply this approach to two complementary problems.  First, we investigate a homogeneous neutrino gas that exhibits a CFI without a fast instability. Representing the neutrino flavor coherence in terms of polarization vectors, the fast dynamics is a simple precession of the polarization vectors around the common flavor axis, so the amplitudes and phases of this precession are left invariant by the fast dynamics. The slow change of these quantities induced by collisions is described by a set of SDEs. We verify that the SDEs lead to the same evolution as the standard EOMs, and we use them to conclude that after the instability has saturated, flavor equipartition is reached for low energies while no conversion is attained for high energies.

Second, we consider a homogeneous neutrino gas that possesses a fast instability, subject to energy-independent collisions. In this case, the fast dynamics, without collisions, is pendular in nature~\cite{Padilla-Gay:2021haz,Fiorillo:2023mze}. Therefore, the slowly-varying quantities can be chosen as the pendulum parameters. In this case, the SDEs allow one to understand immediately several features that had previously only been recovered empirically, such as the observation that the final outcome of the oscillation only depends on the depth of the pendulum and not on any other parameter or the collision rate~\cite{Padilla-Gay:2021haz}.

The structure of the paper follows three main subjects. In Sect.~\ref{sec:active_sterile}, we consider a toy system consisting of an active and a sterile neutrino species subject to oscillations and collisions. This system has been thoroughly studied in the literature and we use it as a simple introductory example for fast dynamics, driven by oscillations, that can be cleanly separated from a slow dynamics, driven by collisions. In Sect.~\ref{sec:cfi}, we discuss the SDEs for a homogeneous system that exhibits CFIs. In Sect.~\ref{sec:pendulum}, we discuss the SDEs for a homogeneous system subject to a fast instability. Finally, in Sec.~\ref{sec:conclusions}, we give a general discussion of our method and its limitations.

\section{Active-sterile neutrino conversion}\label{sec:active_sterile}

The simultaneous appearance of fast and slow dynamics is more general than phenomena in collective neutrino oscillations. A first example for such a separation of time scales is the production of sterile neutrinos $\nu_s$ from a population of active ones $\nu_a$ in the early Universe \cite{Manohar:1986gj, Barbieri:1989ti, Cline:1991zb, Enqvist:1991qj, Dodelson:1993je, Abazajian:2001nj, Hannestad:2012ky, Abazajian:2012ys} or in SN cores \cite{Kainulainen:1990bn, Raffelt:1992bs, Shi:1993ee, Nunokawa:1997ct, Hidaka:2007se, Raffelt:2011nc, Arguelles:2016uwb, Suliga:2019bsq, Suliga:2020vpz}. Even in the standard cosmological model, averaging over the rapid oscillations has proven a good strategy to simplify the numerical solution of the flavor evolution of the neutrino plasma~\cite{Froustey:2020mcq,Froustey:2021azz}. For illustration, we use the simplest toy model, a homogeneous gas of $\nu_a$ and $\nu_s$, no cosmic expansion, and only $\nu_a$ interacting with the ambient plasma. The $\nu_s$ production rate is considerably simplified by the oscillation rate being much faster than the $\nu_a$ collision rate. This simplification is analogous to the slow-dynamics approximation that we will use in the later collisional and fast instabilities examples.

The neutrino radiation field of this simple two-flavor system is described, in the mean-field approximation, by a $2\times2$ density matrix $\rho_\bp(t)$ for every momentum mode $\bp$. Assuming also isotropy, it depends only on $p=|\bp|$ and we write it in the form
\begin{equation}
    \rho_p = 
    \begin{pmatrix}
    f_{a}(p,t) & f_{as}(p,t) \\
    f_{sa}(p,t) & f_{s}(p,t)
    \end{pmatrix},
\end{equation}
where the time dependence of $\rho_p$ is not explicitly shown. The entries are the usual occupation numbers, defined as \smash{$f_{\alpha\beta}(\bp)=\langle a_{\beta,\bp}^\dagger a_{\alpha,\bp}\rangle$} with $\alpha,\beta=a$ or $s$, i.e., the expectation values of number operators, where the mixed off-diagonal elements encode flavor coherence. For the diagonal elements, we do not show the repeated flavor index and we note that the matrix is Hermitian with $f_{sa}(p,t)=f^*_{as}(p,t)$.

In the absence of collision, $\rho_p$ evolves by flavor oscillations caused by masses and flavor mixing and modified by refractive effects of the ambient medium. Without spelling out the details of the latter, we simply assume for every $p$ a Hamiltonian $2\times2$ matrix $\sH_p^0$ for every mode that drives flavor oscillations according to
\begin{equation}
    \dot\rho_p=i\left[\rho_p,\sH_p^0\right].
\end{equation}
(We use sans-serif letters such as $\sH$ to denote matrices in flavor space.)  We assume that $\sH_p^0$ does not depend on the neutrino medium itself so that this EOM is indeed linear in $\rho_p$. In this case, it is explicitly solved by
\begin{equation}\label{eq:simplesolution}
    \rho_p(t)=e^{-i\sH_p^0 t}\rho_p(0)e^{i\sH_p^0 t},
\end{equation}
which can be spelled out analytically because $\sH_p^0$ itself does not depend on time.

The eigenstates 1 and 2 of $\sH_p^0$ are the propagation eigenstates, which for simplicity we call mass eigenstates, with the energies $E_{1,2}=(p^2+m_{1,2}^2)^{1/2}-p\simeq m_{1,2}^2/2p$, where the effective masses depend on $p$, the relativistic approximation was taken, and we have removed the common $p$ from the definition of $\sH_p^0$. The diagonalization of $\sH_p^0$
is enacted by the in-medium mixing angle $\theta$ through
\begin{equation}
    \frac{1}{2p}\begin{pmatrix}m_1^2&0\\0&m_2^2\end{pmatrix}=
    \begin{pmatrix}\cos\theta&\sin\theta\\-\sin\theta&\cos\theta\end{pmatrix}
    \sH_p^0
    \begin{pmatrix}\cos\theta&-\sin\theta\\\sin\theta&\cos\theta\end{pmatrix},
\end{equation}
where it is understood that $\theta$ and $m_{1,2}$ depend on $p$.

In addition, the active states interact with the medium, breaking $\nu_a$--$\nu_s$ flavor coherence. We model this effect by a scattering rate $\Gamma_{p\to p'}$ between $\nu_a$ states and denote with \smash{$\Gamma_p=\sum_{p'}\Gamma_{p\to p'}$} the total loss rate. With the matrix
\begin{equation}\label{eq:Bmatrix}
    \sB=\begin{pmatrix}1&0\\0&0\end{pmatrix}=\frac{1+\sigma_z}{2},
\end{equation}
the modified EOMs are~\cite{Dolgov:1980cq}
\begin{equation}
    \dot{\rho}_p=i\left[\rho_p,\sH^0_p\right]+\sum_{p'}\Gamma_{p'\to p} \sB \rho_{p'} \sB-\frac{\Gamma_p}{2}\left\{\sB,\rho_p\right\}, 
\end{equation}
where we neglect Pauli blocking factors.

The separation of fast and slow timescales naturally happens if the scattering rates are much slower than the oscillations induced by $\sH^0_p$; more precisely, we require that $\Gamma_{p\to p'}$ is much smaller than the difference between the eigenvalues of $\sH^0_p$, i.e., $\Delta m^2/2p\gg\Gamma_p$. In this case, neutrinos oscillate many times between collisions and we can significantly simplify the equations. Based on the no-collision solution of Eq.~\eqref{eq:simplesolution} we first pass to the interaction representation by
\begin{equation}
    \varrho_p(t)=e^{i\sH^0_p t} \rho_p(t) e^{-i\sH^0_p t},
\end{equation}
where the density matrix $\varrho_p$ in the interaction representation would be constant in the absence of collisions. Including collisions, the EOM becomes
\begin{widetext}
    \begin{equation}\label{eq:eom_active_sterile}
    \dot{\varrho}_p=\sum_{p'}\Gamma_{p'\to p} e^{i\sH^0_pt}\,\sB\, e^{-i\sH^0_{p'}t}\varrho_{p'}e^{i\sH^0_{p'}t}\,\sB\,e^{-i\sH^0_p t} 
    -\,\frac{\Gamma_p}{2}\left\{e^{i\sH^0_p t}\sB\,e^{-i\sH^0_p t}, \varrho_p\right\}.
\end{equation}
We can now perform an average over many periods of the rapidly oscillating exponentials. In the mass basis, this means that the off-diagonal entries of any matrix average to zero. For the density matrices, this means explicitly
\begin{equation}
    \varrho_p=\begin{pmatrix} f_1(p) & 0 \\
    0 & f_2(p)
    \end{pmatrix}, \
\end{equation}
where $f_1(p)$ and $f_2(p)$ are the distribution function of the two mass eigenstates. After inserting this parameterization in the EOM~\eqref{eq:eom_active_sterile}, we obtain finally
\begin{subequations}\label{eq:active_sterile_mass}
\begin{eqnarray}
    \dot{f}_1(p)&=&-\Gamma_p f_1(p) \cos^2\theta 
    +\sum_{p'}\Gamma_{p'\to p}
    \left[f_1(p')\cos^4\theta+f_2(p')\sin^2\theta\cos^2\theta\right],
    \\
    \dot{f}_2(p)&=&-\Gamma_p f_2(p)\sin^2\theta
    \,+\sum_{p'}\Gamma_{p'\to p}\left[f_1(p')\sin^2\theta\cos^2\theta
    +f_2(E')\sin^4\theta\right].
\end{eqnarray} 
\end{subequations}
\end{widetext}
These equations refer to the mass-basis distribution functions, which coincide with the flavor-basis distributions if the mixing angle is small. In this case, if the $\nu_a$ are in equilibrium, one finds that the $\nu_s$ production rate equals the $\nu_a$ scattering rate times a factor $\sin^2\theta\cos^2\theta=\sin^2(2\theta)/4$. True equilibrium, however, can only obtain in the mass basis.

The main feature of this formulation is that the right-hand side of Eqs.~\eqref{eq:active_sterile_mass} only contains the scale set by the collision rate. The terms depending on the rapid oscillations have disappeared after we have averaged them out over many periods of oscillation. In the remainder of this paper, we will apply the same strategy to the much more complicated case in which the rapid oscillations are induced by neutrino-neutrino refraction.

\section{Collisional instability}\label{sec:cfi}

\subsection{Setup of the problem}

As a first nontrivial example of the slow/fast dynamics separation, we consider a homogeneous neutrino gas without a fast instability which exhibits CFIs with energy-dependent collisions. We consider the limit of vanishing neutrino masses, i.e., $\omega=\Delta m^2/2p\rightarrow 0$, which are not required for driving the CFI dynamics. We limit ourselves to an isotropic, two-flavor setting with neutrinos subject to mutual forward scattering and number-changing processes with an external medium. In the limit of vanishing neutrino masses and for a homogeneous and isotropic system, the refractive matter term is eliminated by a uniform rotation around the flavor axis. Moreover, the $\Delta m^2\to0$ limit also implies the absence of vacuum mixing. Of course, we still imagine that the mass term provides an initial seed for the CFI.

As in Sect.~\ref{sec:active_sterile}, the system is described by density matrices $\rho_E(t)$ for the flavors $e$ and $x$, where the latter stands for $\mu$ or $\tau$ or an admixture of both. We here use $E\simeq|\bp|$ to label the modes and in addition, we use a negative sign for $E$ to denote the density matrices for antineutrinos. On the other hand, we {\em do not use the flavor isospin convention}, i.e., the upper left entry of $\rho_E$ is the $\nu_e$ or $\overline\nu_e$ occupation number. In the language of polarization vectors to be used later, this convention means that a given $\bP_E$ pointing in the positive $z$-direction means $\nu_e$ or $\overline\nu_e$, depending on the sign of $E$, and pointing down the other flavor. The evolution of the system in the absence of collision is now ruled by the equation
\begin{equation}
    \dot{\rho}_E=i\sqrt{2}G_{\rm F}\int_{-\infty}^{+\infty} \frac{E^{'2}dE'}{2\pi^2}[\rho_E,s_{E'}\rho_{E'}],
\end{equation}
where $s_{E'}={\rm sign}\,E'$ needs to be explicitly included. In order to avoid cluttering notation, and to connect with the definition of a typical energy scale, we define $\mu=\sqrt{2}G_{\rm F} n_\nu^0$, where $n_\nu^0=3T^3\zeta_3/4\pi^2$ is the equilibrium number density of a neutrino population at the temperature $T$ of the thermal bath. We then define a sum over energies with a modified measure $\sum_E=\int E^2 dE/(2\pi^2 n_\nu^0)$. Therefore, the density matrix for the full ensemble is
\begin{equation}
    \rho(t)=\sum_E\,s_E\,\rho_E(t).
\end{equation}
With these definitions, in the absence of collisions, the EOM for every mode has the usual form $\dot\rho_E=i\mu[\rho_E,\rho]$; $\mu$ is a typical energy scale associated with the self-interaction, while the sum over energies, as well as $\rho_E$, is now dimensionless. Notice that, as defined here, $\mu$ does not fully coincide with the usual definition in terms of the total neutrino number density $n_\nu$ (see Sec.~\ref{sec:introduction}); in the context of CFIs such a definition cannot be adopted, since the number of neutrinos is not conserved.

To model the collisional effect, we envision a SN medium where $\nu_e$ and $\overline\nu_e$ are efficiently absorbed or emitted by beta processes, whereas the $x$ flavor does not interact at all. In this sense, our setup is similar to the active-sterile system of Sect.~\ref{sec:active_sterile}. The EOM with beta processes included can be written\footnote{Erratum: In the transition from Eq.~(13) to (16) a factor of 2 was lost, i.e., in
Eq.~(16) and following one should substitute $Q_E\to Q_E/2$ and 
$\Gamma_E\to \Gamma_E/2$. Alternatively, the definition in Eq.~(13)
should be changed to $Q_E {\sf B}=2{\sf P}_E$ and  $\Gamma_E {\sf
B}=2({\sf A}_E+{\sf P}_E)$. Overall conclusions remain unchanged.} in the form~\cite{Sigl:1993ctk}
\begin{eqnarray}\label{eq:eom_original}
    \dot\rho_E&=&i\mu[\rho_E,\rho]+\frac{1}{2}\left\{\sP_E,1-\rho_E\right\}
    -\frac{1}{2}\left\{\sA_E,\rho_E\right\}
\nonumber\\
  &=&i\mu[\rho_E,\rho]+{\underbrace{\sP_E}_{\textstyle Q_E\sB}}
  -\frac{1}{2}\bigl\{\underbrace{\sA_E+\sP_E}_{\textstyle \Gamma_E\sB}
  ,\rho_E\bigr\},
\end{eqnarray}
where in the flavor basis the production and absorption matrices $\sP_E$ and $\sA_E$ have only upper-left entries like $\sB$ in Eq.~\eqref{eq:Bmatrix}, relevant for $\nu_e$ and $\overline\nu_e$. 

Here, $Q_E$ is the spontaneous emission rate of $\nu_e$ by the medium, $\Gamma_E$ the reduced absorption rate, and analogous for $\overline\nu_e$. Ignoring temporarily the matrix structure and considering only the $e$-flavor, the collision equation is $\dot f_e=Q_E(1-f_e)-\Gamma_E^0 f_e$, where $\Gamma_E^0$ is the absorption rate. Collecting the terms linear in $f_e$, this is $\dot f_e=Q_E-\Gamma_E f_e$ with $\Gamma_E=\Gamma_E^0+Q_E$ the ``reduced'' absorption rate, although it is actually enhanced, the terminology deriving from photon radiative transport, where the Bose stimulation factor has the opposite sign from the Pauli blocking factor. Moreover, if the medium is in thermal equilibrium, detailed balance implies for~$\nu_e$ (positive $E$) $Q_E=\Gamma^0_E e^{-(E-\mu_{\nu_e})/T}$ and thus $\Gamma_E=\Gamma^0_E[1+e^{-(E-\mu_{\nu_e})/T}]$, where $\mu_{\nu_e}$ is the $\nu_e$ chemical potential implied by the properties of the medium. Using $\Gamma_E$ as our primary parameter for the damping strength, the production rate is
\begin{equation}\label{eq:QEGE}
    Q_E=\frac{\Gamma_E}{e^{\pm(E-\mu_{\nu_e})/T}+1},
\end{equation}
where $\pm$ refers to $\nu_e$ and $\overline\nu_e$. In our convention, the latter are denoted by negative $E$ and have the opposite chemical potential. Also, for negative $E$, $\Gamma_E$ is the reduced absorption rate for antineutrinos.

The equations are most conveniently expressed in terms of the total number $P^0_E = \mathrm{Tr}\,\rho_E=f_e(E)+f_x(E)$ of neutrinos in mode $E$ and the polarization vectors $\bP_E$, such that
\begin{equation}
    \rho_E=\frac{P^0_E}{2}+\frac{\bP_E\cdot \boldsymbol{\sigma}}{2}.
\end{equation}
In terms of these variables, the EOMs become
\begin{subequations}
\begin{eqnarray}\label{eq:eom_1a}
    \kern-2em \dot{\bP}_E&=&\mu \bP\times\bP_E - \Gamma_E \bP_E +(2Q_E-\Gamma_E P^0_E)\,\be_z,
    \\[1ex] \label{eq:eom_2a}
    \kern-2em \dot{P}^0_E&=&2Q_E-\Gamma_E P^0_E-\Gamma_E P^z_{E}.
\end{eqnarray}    
\end{subequations}
Here $\be_z$ identifies the direction of the flavor axis, and $\bP=\sum_E s_E \bP_E$. 

It will prove convenient to express them in terms of a modified neutrino density $N_E=P^0_E-2Q_E/\Gamma_E$, a quantity that need not be positive. In thermal and chemical equilibrium, both $e$ and $x$ flavors are occupied with the Fermi-Dirac distribution $f_{e,x}(E)=1/[e^{(E-\mu_{\nu_e})/T}+1]$ and $P_E^0= f_{e}(E)+f_{x}(E)$ with Eq.~\eqref{eq:QEGE} reveals that $N_E$ is the deviation from equilibrium. Finally, 
\begin{subequations}
\begin{eqnarray}\label{eq:eom_1}
    \dot{\bP}_E&=&\mu \bP\times\bP_E - \Gamma_E (\bP_E + N_E\be_z),
    \\[1ex]
    \dot{N}_E&=&-\Gamma_E (N_E+P^z_E)
\end{eqnarray}    
\end{subequations}
is the set of EOMs that we plan to solve.

\subsection{Slow-dynamics equations (SDEs)}

We now proceed to integrate out the fast dynamics in the spirit of our earlier discussion. When $\Gamma_E=0$, the exact solution consists of a fast precession of each $\bP_E$ around $\bP$. This motion is simple to understand and has many conserved quantities. In the absence of collisions, the total $\bP=\sum_E s_E\bP_E$ is conserved, reflecting total lepton-number conservation. Furthermore, each $\bP_E$ moves uniformly around $\bP$ on a cone with a fixed angle. To write the general motion explicitly, we introduce a right-handed coordinate system $\{\be_1,\be_2,\be_3\}$, where $\be_3$ is along $\bP$, whereas $\be_1=\be_z\times\bP/|\be_z\times\bP|$ is orthogonal to both $\bP$ and the flavor direction $\be_z$, and finally $\be_2=\be_3\times\be_1$. In other words, $\be_1$ and $\be_2$ span the plane transverse to~$\bP$. The projection of $\bP_E$ on the direction of $\bP$ is expressed through $\alpha_E=\bP_E\cdot\be_3/P$, whereas the length in the transverse plane is $b_E=|\bP_E\times\be_3|$. In this way, the explicit solution~is
\begin{equation}\label{eq:solution}
\bP_E(t)=\alpha_E \bP + b_E \bigl[\cos\left(\Phi_t +\phi_E\right)\be_1
+\sin\left(\Phi_t + \phi_E\right)\be_2\bigr],
\end{equation}
where the $\phi_E$ are fixed individual phases and $\Phi_t=\mu P t$ the common oscillation phase. When we later assume that $\bP$ slowly changes, it will be \smash{$\Phi_t=\mu\int_0^t dt'\,P(t')$}. The solution is entirely specified by the conserved quantities $\bP$, the projection $\alpha_E$ of each $\bP_E$ on the precession axis, the amplitude of precession $b_E$, and the phase $\phi_E$. Because $\bP=\sum_E s_E\bP_E(t)$, these quantities are constrained by $\sum_Es_E\alpha_E=1$, $\sum_Es_Eb_E \cos\phi_E=0$,
and $\sum_E  s_Eb_E \sin \phi_E=0$. 

Collisions cause these quantities to slowly evolve over timescales given by the inverse damping rate. By the assumption of weak damping, each $\bP_E$ performs many precession cycles over this timescale and therefore the precession itself can be averaged out. In analogy to Sect.~\ref{sec:active_sterile}, we replace the form of the solution Eq.~\eqref{eq:solution} in the full EOMs with damping, where we now consider the parameters $\bP$, $\alpha_E$, $b_E$, $\be_1$, $\be_2$, and $\phi_E$ as dynamical quantities. Notice that the directions $\be_1$, $\be_2$ and $\bP$ themselves depend on time and therefore must be differentiated as well. To average over the rapidly varying phase $\Phi_t$, we use $\langle \sin \Phi_t\rangle=\langle\cos \Phi_t \rangle=0$, $\langle \cos\left(\Phi_t+\alpha\right)\cos\left(\Phi_t+\beta\right)\rangle=\langle \sin\left(\Phi_t+\alpha\right)\sin\left(\Phi_t+\beta\right)\rangle=\frac{1}{2}\cos(\alpha-\beta)$, and the mixed $\langle \cos\left(\Phi_t+\alpha\right)\sin\left(\Phi_t+\beta\right)\rangle=\frac{1}{2}\sin(\beta-\alpha)$. Performing these averages directly provides us with the SDEs:
\begin{widetext}
\begin{subequations}
    \begin{eqnarray}\label{eq:proj_adia}
        \dot{\alpha}_E&=&\alpha_E \sum_{E'} s_{E'} \Gamma_{E'}
        \left(\alpha_{E'}+\frac{N_{E'} P^z}{P^2}\right)
       -\Gamma_E\left(\alpha_E+\frac{N_E P^z}{P^2}\right)
    -\frac{b_E}{P^2}\sum_{E'} s_{E'} \Gamma_{E'} b_{E'} \cos(\phi_E-\phi_{E'}),
        \\ \label{eq:ampl_adia}
        \dot{b}_E&=&\alpha_E \sum_{E'} s_{E'} \Gamma_{E'} b_{E'} \cos(\phi_E-\phi_{E'})-\Gamma_E b_E,
        \\ \label{eq:phase_adia}
        b_E \dot{\phi}_E&=&-\alpha_E \sum_{E'} s_{E'} \Gamma_{E'} b_{E'} \sin(\phi_E-\phi_{E'})
        \\ \label{eq:pol_adia}
        \dot{\bP}&=&-\sum_{E'} s_{E'} \Gamma_{E'}
        \bigl(N_{E'} \be_z+\alpha_{E'} \bP\bigr),
        \\ \label{eq:ndensity_adia}
        \dot{N}_E&=&-\Gamma_E N_E -\Gamma_E \alpha_E P^z.
    \end{eqnarray}
\end{subequations}    
\end{widetext}
These equations are admittedly not much simpler than the original EOMs; they are still nonlinear and couple neutrinos of all energies. However, they offer some practical and conceptual advantages. There is no term of order $\mu$ and thus a numerical solution requires only a relatively coarser grid, with a step of order $\Gamma^{-1}$ that mirrors the scale over which the quantities change. More importantly, the new equations provide some intuitive understanding of the role of damping, by cleaning the dynamics of the complicated precession terms.

As a first step, we can eliminate the phase variables $\phi_E$. From Eq.~\eqref{eq:phase_adia} we see that the phases follow a Kuramoto-like dynamics~\cite{Kuramoto1975, Kuramoto2003, Pantaleone:1998xi}, with the damping term tending to synchronize the phases of precession and make all the polarization vectors coplanar. If two vectors $\bP_E$ and $\bP_{E'}$ have identical phases $\phi_E=\phi_{E'}$, one can see from Eq.~\eqref{eq:solution} that at a given time they lie in the plane spanned by $\bP$ and $\cos(\Phi_t+\phi_E)\be_1+\sin(\Phi_t+\phi_E)\be_2$. (Notice that for a two-beam case the two vectors are necessarily co-planar, so one can exactly set $\phi=0$.) One can always choose the initial condition such that all $\bP_E$ are exactly coplanar, and the subsequent dynamics likely does not depend on this choice, since anyway they would soon become coplanar because of phase locking. In the following, we simply choose $\phi_E=0$ for every $E$.

We can now understand the existence of two separate branches of CFIs from the SDEs directly. Starting from Eq.~\eqref{eq:pol_adia}, let us first consider the component transverse to the flavor direction,
\begin{equation}
\dot{\bP}_T=-\sum_E s_E \Gamma_E \alpha_E \bP_T.
\end{equation}
Thus, the evolution of the transverse component is driven by the effective damping rate $\sum_E s_E \Gamma_E \alpha_E$. If initially this expression is negative, it becomes an effective growth rate, leading to an instability and coincides with one of the two branches of instability that have been identified in the literature~\cite{Johns:2021qby, Xiong:2022zqz, Fiorillo:2024}.

Even when this branch corresponds to damping, an instability can still arise from the growth of the individual precession amplitudes $b_E$. To see how this can happen, we take the simplest case of a two-bin model with $\overline{b}=b$ and $\overline{\alpha}=\alpha-1$, where we denote by $\alpha$, $b$, and $\Gamma$ the quantities for monochromatic neutrinos, and by $\overline{\alpha}$, $\overline{b}$, and $\overline{\Gamma}$ the ones for antineutrinos. Equation~\eqref{eq:ampl_adia} then simplifies to
$\dot{b}=\alpha(\Gamma-\overline{\Gamma})b-\Gamma b$ that can be written~as
\begin{equation}\label{eq:secondinstability}
\dot{b}=\left(\alpha\overline{\Gamma}-\Gamma \overline{\alpha}\right)b.
\end{equation}
If the expression in brackets is positive, $b$ grows, corresponding to the second branch of instability identified in the previous literature. In the two-bin model, the first branch corresponds to $(\overline\alpha\overline{\Gamma}-\Gamma {\alpha})$ being positive (growth) or negative (damping). Thus, for this simple two-bin system, we recover the linear stability results for the isotropic, non-resonant collisional instability~\cite{Lin:2022dek,Liu:2023pjw}.

The identification of the instability allows one to provide a clearer criterion for the applicability of the SDE approach. So far, we have schematically required a hierarchy of the form $\mu \gg \Gamma$, but since the interaction rate $\Gamma_E$ is actually energy-dependent, this requirement should be made more specific. The typical timescale over which the CFI evolves is indicated by the collective growth rate, which we denote by $\gamma$, determined by linear stability analysis.  On the other hand, the precession of the polarization vectors happens on typical timescales of the order of $\mu P^z(t=0)$. Therefore, a more concrete criterion for the applicability of the SDEs is that $\mu P^z(t=0)\gg \gamma$.

\subsection{Understanding the nonlinear outcome}

\subsubsection{Simplified equations of motion}

The second instability, in the two-bin case represented by Eq.~\eqref{eq:secondinstability}, corresponds to the presence of more $\nu_e$ than $\overline\nu_e$, and $\nu_e$ more strongly damped, which is the situation in a SN core with regard to beta processes. Therefore, we focus on this case and gather some physical intuition about its impact on the flavor evolution. This case corresponds to damping for the first type of mode, i.e., damping in Eq.~\eqref{eq:pol_adia} so that $\bP$ never develops a large transverse component if we begin with all $\bP_E$ aligned with the $z$-direction except for a small seed. Therefore, we can simplify our equations by taking $P=P^z$.

We can further simplify the EOMs by writing them in terms of dimensionless parameters through $N_E=\nu_E P$ and $b_E=\beta_E P$. The interpretation of the new parameters is that for every $E$, $\alpha_E$ is the dimensionless $z$ component of $\bP_E$, $\beta_E$ the dimensionless amplitude of precession, and $\nu_E$ represents the dimensionless number of neutrinos in that mode. We further define
\begin{subequations}
    \begin{eqnarray}\label{eq:Adefinition}
        A&=&\sum_E s_E \Gamma_E \alpha_E,
    \\ \label{eq:Bdefinition}
        B&=&\sum_E s_E \Gamma_E \beta_E,
    \\ \label{eq:Ndefinition}
        N&=&\sum_E s_E \Gamma_E \nu_E.
    \end{eqnarray}
\end{subequations}
The EOMs of the dimensionless variables are then
\begin{subequations}\label{eq:adia_full_system}
\begin{eqnarray}
\dot{\alpha}_E &=& (N+A-\Gamma_E)\alpha_E-\Gamma_E\nu_E+B\beta_E,\\  
\dot{\beta}_E &=&  (N+A-\Gamma_E)\beta_E+B\alpha_E,\\
\dot{\nu}_E &=& (N+A-\Gamma_E)\nu_E-\Gamma_E \alpha_E,
\end{eqnarray}    
\end{subequations}
whereas
\begin{equation}
\dot{P}=-(N+A)P
\end{equation}
regulates the dynamics of $P$.

For the absorption rates we assume a quadratic energy dependence $\Gamma^0_E=\Gamma (E/3T)^2$ for $\nu_e$ (positive $E$) and $\overline{\Gamma}(E/3T)^2$ for $\overline\nu_e$ (negative $E$), where $\Gamma$ and $\overline\Gamma$ are positive parameters. As explained in the text above Eq.~\eqref{eq:QEGE}, the reduced damping parameters then have the form
\begin{subequations}
    \begin{eqnarray}
     \kern-3em   \Gamma_E&=&\Gamma \left(\frac{E}{3T}\right)^2 \left[1+e^{-(E-\mu_{\nu_e})/T}\right]
        \quad\hbox{for $E>0$},
        \\
      \kern-3em   \Gamma_E&=&\overline{\Gamma}\left(\frac{E}{3T}\right)^2 \left[1+e^{+(E-\mu_{\nu_e})/T}\right]
        \quad\hbox{for $E<0$},
    \end{eqnarray}
\end{subequations}
where $\mu_{\nu_e}$ is the $\nu_e$ chemical potential defined by the medium properties. Here $\Gamma$ and $\overline\Gamma$ are the damping rates for typical $\nu_e$ or $\overline\nu_e$ energies.

Our simplified EOMs immediately reveal that the final state only depends on the ratio $\overline{\Gamma}/\Gamma$ because all damping parameters are linear in $\Gamma_E$ so that $\Gamma$ can be pulled out and absorbed in the definition of the units of time.

\subsubsection{Numerical example}

To build some intuition, we now turn to a specific example, for which we solve numerically the newly derived SDEs~\ref{eq:adia_full_system}. As initial distributions, we choose thermal ones for $\nu_e$ and $\overline{\nu}_e$
\begin{equation}\label{eq:thermaldistribution}
    \frac{dn_\nu}{d\epsilon}=\frac{1}{2\pi^2}\,
    \frac{\epsilon^2}{e^{(\epsilon-\mu_\nu)/T}+1},
\end{equation}
corresponding to the distribution of beta equilibrium with the medium. Here we use $\epsilon=|E|$ as a positive energy variable. For $\nu_x$ and $\overline{\nu}_x$, we assume that their initial population essentially vanishes, so they can only be produced by flavor conversion. The numerical values chosen for the parameters are listed in Table~\ref{tab:params}. We also show the initial number densities that follow from Eq.~\eqref{eq:thermaldistribution}. The number density of neutrinos with vanishing chemical potential for $T=4$~MeV is $n_\nu^0=5.864~{\rm MeV}^3$. With our definition of the sum over energies, the initial value of the total polarization vector is $P^z(0)=\frac{1}{2}(n_{\nu_e}-n_{\overline{\nu}_e})/n_\nu^0=0.468$.

\begin{table}[ht]
\caption{Parameters used for the numerical simulation.\label{tab:params}}
\vskip4pt
\begin{tabular*}{\columnwidth}{@{\extracolsep{\fill}}lllll}
\hline
Species & $T$ [MeV] & $\mu_\nu$ [MeV]&$n_\nu$ [MeV$^{3}$]&$n_\nu/n_\nu^0$\\
\hline
$\nu_e$ & 4 & ~~2        &9.146&1.564\\
$\overline{\nu}_e$ & 4 & $-2$ &3.678&0.629\\
\hline
\end{tabular*}
\end{table}

As discussed earlier, the absolute values of $\Gamma$ and $\overline{\Gamma}$ are irrelevant, and only their relative values to one another matter, as long as we are in the weak-damping regime $\Gamma,\overline{\Gamma}\ll\mu$. We choose the ratio $\overline{\Gamma}/\Gamma=0.1$ and measure time in units of $\Gamma^{-1}$. 

We initialize the system with seeds of small randomly chosen off-diagonal components $P^x_E$ and $P^y_E$ ($|P^x_E,P^y_E|\lesssim 10^{-3}P^z_E$ for all energies) . As a reference comparison, we solve the SDEs and compare them with the solution of the full EOMs for $\mu=50\,\Gamma$. Linear stability analysis shows that the growth rate of the CFI is $\gamma\simeq 0.39 \Gamma$, so that $\mu P^z(0)\simeq 122\,\gamma$, justifying the applicability of the SDE approach.

\begin{figure*}[ht]
    \centering
    \includegraphics[width=0.90\textwidth]{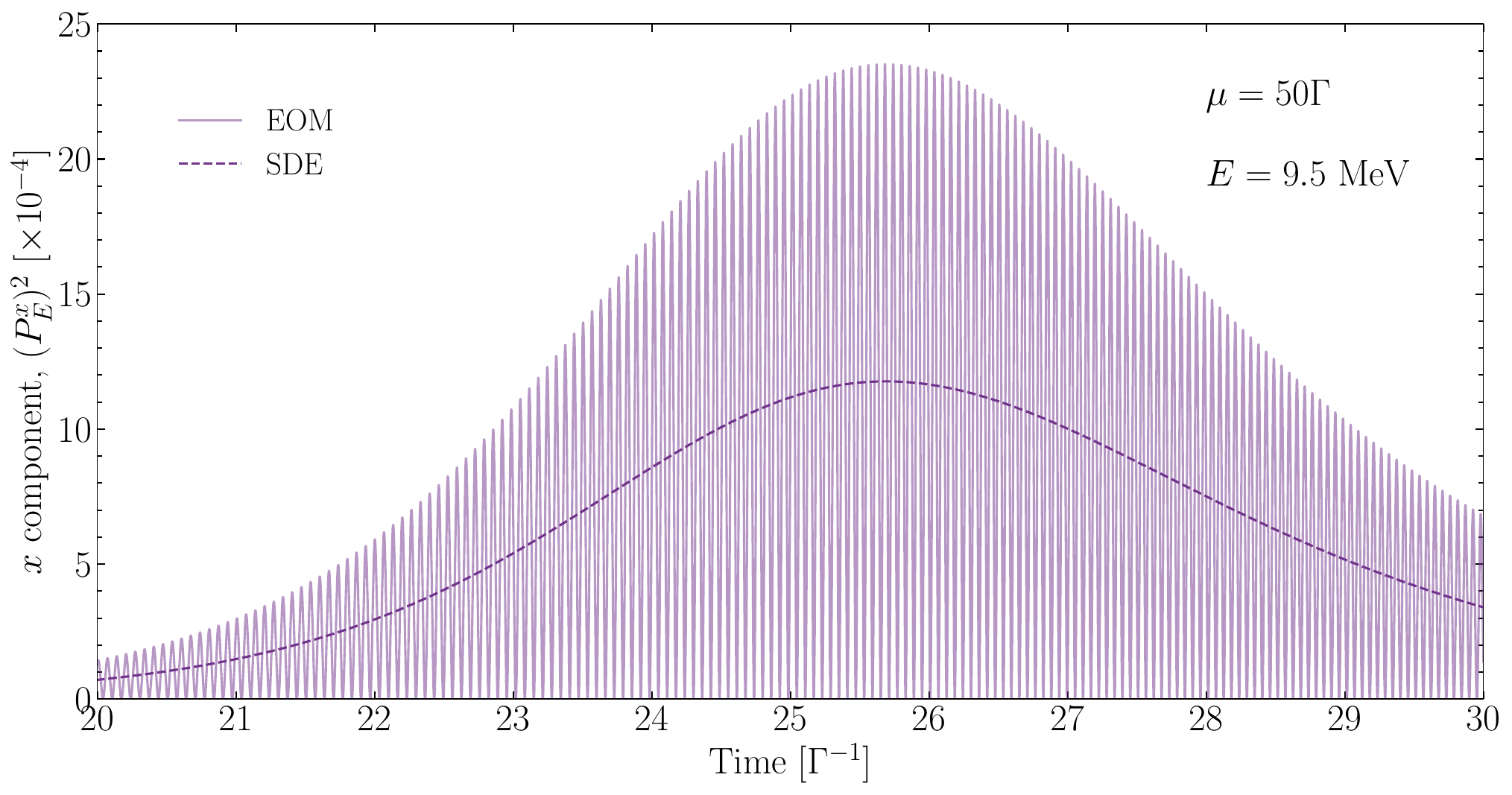}
    \caption{Comparison of the solution of the full EOMs (solid thin) and the solution of the SDEs (dashed thick) for the squared transverse component of the polarization vector corresponding to an energy $E=9.5$~MeV.}
    \label{fig:fast_vs_slow}
\vskip24pt
    \includegraphics[width=\textwidth]{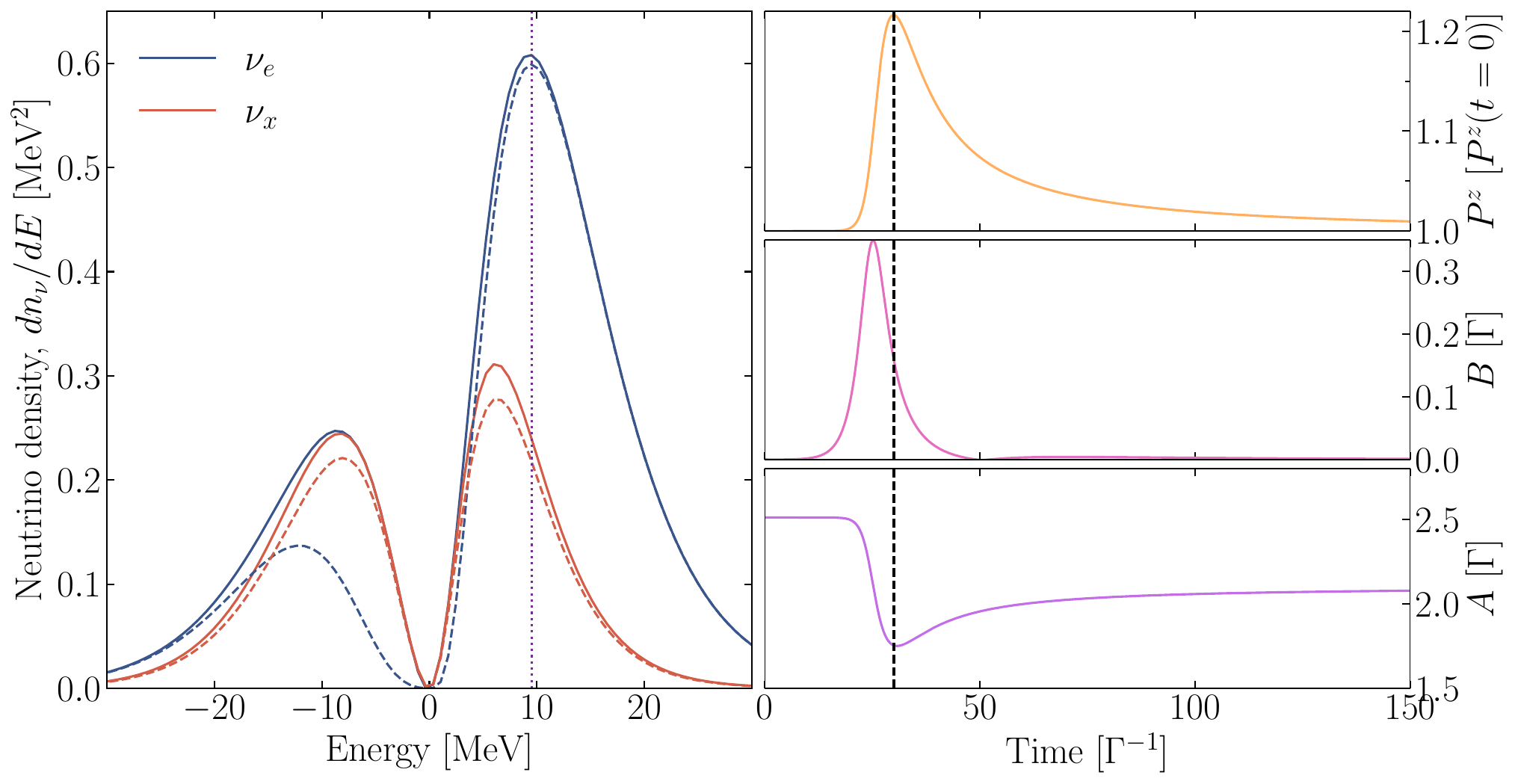}
    \caption{Numerical solution of the SDEs for the model system described in the text. \textit{Left:} Final (solid) and intermediate (dashed) spectrum of $\nu_e$ and $\nu_x$ ($\overline{\nu}_e$ and $\overline{\nu}_x$ are shown on the negative energy axis). For $\nu_e$ and $\overline\nu_e$, the final spectrum is identical to the initial one, corresponding to beta equilibrium with the medium. The intermediate spectrum corresponds to the instant in time when $P^z$ is maximum, identified by the dashed line in the right panels. The vertical dotted line identifies the energy of the polarization vector whose dynamics is illustrated in Fig.~\ref{fig:fast_vs_slow}. \textit{Right:} Time evolution of the integrated quantities $P^z$, $B$, and $A$, as defined in the main text. }
    \label{fig:num_simulation}
\end{figure*}

As an example of the simplification induced by the SDEs, we show in Fig.~\ref{fig:fast_vs_slow} the evolution of the squared transverse component $(P^x_E)^2$ of a specific polarization vector, corresponding to an energy $E=9.5$~MeV, where the $\nu_e$ spectrum peaks, both solving the full EOMs and the SDEs. For the latter, from Eq.~\eqref{eq:solution} we see that, since $\bP$ remains aligned with the $z$ axis, the mean squared component is simply $\langle(P^x_E)^2\rangle\simeq b_E^2/2$. This has a simple interpretation; since $b_E$ is the amplitude of the precession around the $z$ axis, the squared amplitude along a specific direction is simply one-half of the transverse squared amplitude. The comparison in Fig.~\ref{fig:fast_vs_slow} shows that the solution of the SDEs tracks the average of the exact solution of the EOMs over the very rapid oscillations. 

Figure~\ref{fig:num_simulation} shows the evolution of this system, as predicted from the SDEs, as well as the final state reached after the CFI saturates. The final $\nu_e$ and $\overline{\nu}_e$ distributions are identical to the initial ones, both corresponding to beta equilibrium with the medium, but during the evolution they deviate. By assumption, there is no initial population of $\nu_x$ and $\overline{\nu}_x$. The dashed lines show the distributions at the intermediate stage chosen at the instant when $P^z$ is maximal as indicated by the vertical dashed line in the right panels.

The CFI is manifested in the growth of the in-plane components of the polarization vectors, namely of the parameters $\beta_E$ introduced above. Correspondingly, the collective amplitude $B=\sum_\epsilon s_E \Gamma_E \beta_E$ grows exponentially in time. We easily check that the instability for these conditions belongs to the second branch identified above, since the in-plane components $P^x$ and $P^y$ never grow, so the total polarization vector remains aligned with the $z$ axis. Its magnitude initially grows exponentially. After the instability grows nonlinear, both $P^z$ and $B$ return to their original value; in the case of $B$, this means that the polarization vectors return to align with the $z$ axis. We also show the evolution of the function $A$ defined in Eq.~\eqref{eq:Adefinition}; the complementary function $N$ defined in Eq.~\eqref{eq:Ndefinition} remains always essentially identical to $N\simeq -A$. As the instability grows, the $\nu_e$ and $\overline{\nu}_e$ population are temporarily depleted since they convert into $\nu_x$ and $\overline{\nu}_x$, as visible from the dashed lines in Fig.~\ref{fig:num_simulation}.

After the instability has saturated, $\nu_e$ and $\overline{\nu}_e$ return to their initial spectrum. This is only to be expected since the equilibrium state of the $e$ flavor is determined by thermal and chemical equilibrium with the medium and is enforced by the collisional term. On the other hand, $\nu_x$ and $\overline{\nu}_x$ do not interact with the medium, and therefore their final amount is not so trivially understood. The simulation reveals that their final distributions are characterized by an equipartition of $\nu_e$ and $\nu_x$ for low energies. At high energies, the $\nu_x$ remain unpopulated and are not efficiently produced by flavor conversion. 

\subsubsection{Low-energy behavior}

To show that these main features identified in the numerical solution descend naturally from the SDEs, we begin with the low-energy behavior. Beginning with Eq.~\eqref{eq:adia_full_system}, we first perform the transformation
\begin{equation}
    \alpha_E=\tilde{\alpha}_E \exp\left(\int_0^t dt'\left[N(t')+A(t')-\Gamma_E\right]\right),
\end{equation}
and a similar transformation for $\beta_E$ and $\nu_E$. In terms of the new variables, the equations become
\begin{subequations}\label{eq:tildes}
\begin{eqnarray}
    \dot{\tilde{\alpha}}_E&=&- \Gamma_E \tilde{\nu}_E+B \tilde{\beta}_E,\\ \dot{\tilde{\beta}}_E &=&B \tilde{\alpha}_E,\\ 
    \dot{\tilde{\nu}}_E   &=&- \Gamma_E \tilde{\alpha}_E.
\end{eqnarray}    
\end{subequations}
We now notice that the behavior of the system is critically different according to whether the ratio $\Gamma_E/B$ is very large or very small. Since the function $B$ depends on time and is initially very small, clearly for all polarization vectors we have initially $B\ll \Gamma_E$. However, as time progresses, $B$ grows exponentially until a maximum value $B_{\rm max}$; for the numerical example analyzed earlier, $B_{\rm max}\simeq 0.35\,\Gamma$, as seen from the right panel of Fig.~\ref{fig:num_simulation}. 

The coefficients $\Gamma_E$ have a quadratic energy dependence, while the Fermi-Dirac factor does not have a dramatic impact, so we can approximately identify a range of energies $E\lesssim 3T\sqrt{B_\mathrm{max}/\Gamma}$ for neutrinos and $E\lesssim 3T\sqrt{B_\mathrm{max}/\overline{\Gamma}}$ for antineutrinos. For these energies, we may get a qualitative insight by considering the limit $\Gamma_E\rightarrow 0$, where Eqs.~\eqref{eq:tildes} result in $(\dot{\tilde{\alpha}}_E-\dot{\tilde{\beta}}_E)=-B(\tilde{\alpha}_E-\tilde{\beta}_E)$ which implies that $\tilde{\alpha}_E-\tilde{\beta}_E$ decreases exponentially in time as $e^{-\int_0^t B(t') dt'}$. Therefore, in this low-energy range, $\alpha_E$ and $\beta_E$ are maintained approximately equal. Since $\beta_E$ finally reaches the value of zero, after the transverse motion has exhausted, we conclude that in this low-energy range, we must finally have $\alpha_E=0$, i.e., equipartition among the $e$ and $x$ flavor. Our numerical simulation completely verifies this prediction, as seen in Fig.~\ref{fig:num_simulation}. Notice that this allows us to conclude that the maximum energy at which equipartition is reached for neutrinos and antineutrinos, $\tilde{E}_{\nu}$ and $\tilde{E}_{\overline{\nu}}$ respectively, are related by $\tilde{E}_{\nu}/\tilde{E}_{\overline{\nu}}\simeq \sqrt{\overline{\Gamma}/\Gamma}$, a conclusion that is verified by our numerical simulation.

\subsubsection{High-energy behavior}

At high energies, such that $\Gamma_E\gg B_\mathrm{max}$, the qualitative behavior of the solution can be understood for the limit $B\rightarrow 0$, where we find $(\tilde{\alpha}_E^2-\tilde{\nu}_E^2)$ to be constant at all times with $\tilde{\alpha}_E, \tilde{\nu}_E\propto e^{-i\Gamma_E t}$. Moreover, $\beta_E$ never becomes large, and in turn, no large flavor conversion occurs; neutrinos remain pinned to their initial configuration. This prediction is again confirmed by Fig.~\ref{fig:num_simulation}, where we see a clear break to a region where flavor equipartition is not reached, and at sufficiently large energies ($E\gtrsim 20$~MeV), the $\nu_x$ spectrum is hardly populated.

\section{Damped fast flavor pendulum}
\label{sec:pendulum}

\subsection{Setup of the problem}

A somewhat more complex situation arises when the fast dynamics, unperturbed by collisions, is not simply a precession. One obvious example is the fast flavor pendulum~\cite{Johns:2019izj,Johns:2020qsk, Padilla-Gay:2021haz}, which in its simplest manifestation arises in a homogeneous, anisotropic, azimuthally symmetric, and monoenergetic neutrino system. In the absence of collisions, if the angular distribution of the lepton number has crossings obeying a certain criterion \cite{Fiorillo:2023hlk}, the system possesses a fast instability and, due to the presence of an infinite number of integrals of motion~\cite{Johns:2019izj, Fiorillo:2023mze}, exhibits a pendulum-like dynamics.

We now add a collisional damping term which is taken to be equal for neutrinos and antineutrinos, a case that was previously examined numerically~\cite{Padilla-Gay:2022wck}~(see also Ref.~\cite{Johns:2022bmu} for a discussion of the impact of collisions on fast instabilities). We will see that our separation of
fast/slow dynamics applied to this system directly recovers the earlier features. In this case, the polarization vectors $\bP_v$ and $\overline{\bP}_v$ for neutrinos and antineutrinos depend on the velocity $-1<v<1$ along the axis of azimuthal symmetry. The dynamics is entirely driven by the lepton number $\bD_v=\bP_v-\overline{\bP}_v$. If the damping rate is $\Gamma$, the EOMs are
\begin{equation}
    \dot{\mathbf{D}}_v=(\bD_0-v\mathbf{D}_1)\times\mathbf{D}_v-\Gamma \be_z\times(\mathbf{D}_v\times\be_z),
\end{equation}
where $\bD_0=\sum_v \bD_v$ and $\bD_1=\sum_v v \bD_v$ and $\be_z$ the flavor direction. The EOM for $\bD_0$ is
\begin{equation}
    \dot{\mathbf{D}}_0=-\Gamma\,\be_z\times(\mathbf{D}_0\times\be_z),
\end{equation}
corresponding to damping of its component transverse to the flavor direction.

In the absence of collisions, no direction in flavor space is singled out and the conserved $\mathbf{D}_0$, given by initial conditions, defines the $z$-direction. Collisions with the background medium, for example beta processes, introduce a flavor direction and define $\be_z$. If the small seeds chosen to start the motion imply that $\mathbf{D}_0$ is not exactly parallel to $\be_z$, this deviation is quickly damped but otherwise, this small misalignment has no further impact and we may assume that $\mathbf{D}_0$ is conserved, simplifying the EOMs.
In this case we may remove the term $\bD_0\times \bD_v$ in the EOMs by a uniform corotation. After these simplifications, 
\begin{equation}
    \dot{\mathbf{D}}_v=-\mu v\mathbf{D}_1\times\mathbf{D}_v
    -\Gamma\,\be_z\times(\mathbf{D}_v\times\be_z)
\end{equation}
are the EOMs to be solved.

\subsection{Pendulum dynamics}

In the absence of collisions, for a fast-unstable system with a single crossing, the motion of the polarization vectors is periodic and can be expressed in terms of three fundamental vectors only. The dynamics of these three vectors is identical to that of a spherical pendulum~\cite{Padilla-Gay:2021haz,Fiorillo:2023mze}. In other words, we can introduce a fictitious system of three polarization vectors $\bP_\alpha$, with three velocities $v_\alpha$, obeying the EOMs
\begin{equation}
    \dot{\bP}_\alpha=-\mu v_\alpha \bP_1\times \bP_\alpha,
\end{equation}
with $\bP_1=\sum_\alpha v_\alpha \bP_\alpha$. The original polarization vectors can be now expressed in terms of this fictitious system using the connection~\cite{Fiorillo:2023mze}
\begin{equation}
    \bD_i=\sum_\alpha\frac{v_\alpha \bP_\alpha}{v_i-v_\alpha}.
\end{equation}
The two systems have identical dynamics, provided they obey the matching condition
\begin{equation}
    \sum_i v_i \bD_i=\sum_\alpha v_\alpha \bP_\alpha;
\end{equation}
this condition needs only be satisfied at the initial time, and it will be automatically true at all later times.

What is surprising is that, for the special choice of the isotropic collisional term we consider for this work, the equivalence between the full system $\bD_i$ and the three-beam system $\bP_\alpha$ is rigorously true also in the presence of collisions. Indeed, by explicit substitution, we find that if $\bP_\alpha$ obey the EOMs
\begin{equation}
    \dot{\bP}_\alpha=-\mu v_\alpha \bP_1\times \bP_\alpha-\Gamma \be_z\times (\bP_i\times \be_z),
\end{equation}
the associate polarization vectors $\bD_i$ automatically satisfy the correct EOMs. Therefore, without loss of generality we restrict our discussion to a system of three velocity beams only.

The dynamics of three beams in the absence of collisions can always be reduced to a pendulum dynamics. We do this by introducing the variables $\mathbf{D}_0=\sum_\alpha \mathbf{P}_\alpha$, $\mathbf{D}_1=\sum_\alpha v_\alpha \mathbf{P}_\alpha$, $\mathbf{J}=\sum_{\alpha} v_\alpha(s -v_\alpha) \mathbf{P}_\alpha$ with $s=\sum_\alpha v_\alpha$, which obey the EOMs
\begin{subequations}
    \begin{eqnarray}
        \dot{\mathbf{D}}_0&=&-\Gamma\be_z\times(\mathbf{D}_0\times\be_z),
        \\[1ex]
        \dot{\mathbf{D}}_1&=&\mu \mathbf{D}_1\times\mathbf{J}-\Gamma\be_z\times(\mathbf{D}_1\times\be_z),
        \\[1ex]
        \dot{\mathbf{J}}&=&\mu p\mathbf{D}_1\times\mathbf{D}_0-\Gamma\be_z\times(\mathbf{J}\times\be_z),
    \end{eqnarray}
\end{subequations}
with $p=\prod_\alpha v_\alpha$. For $\Gamma=0$, these indeed coincide with the EOMs of a spherical pendulum. 

\subsection{Slow dynamics evolution of flavor pendulum}

The unperturbed pendulum motion, for $\Gamma=0$, is characterized by the invariants of motion $D_1$, $J_z=\mathbf{J}\cdot\be_z$, $\alpha=\mathbf{J}\cdot\mathbf{D}_1/D_1$, and $E=\frac{1}{2}\mathbf{J}^2+p\mathbf{D}_0\cdot\mathbf{D}_1$. Collisions lead to a slow evolution of these quantities. Therefore, just as in the case of CFIs, we can now obtain the SDEs describing only the slow change of these quantities over timescale $\sim \Gamma^{-1}$, rather than the fast pendular motion over timescales $\sim \mu^{-1}$. As in the case of CFI, we need to average the right-hand side of the corresponding equations over many periods of the fast 
motion induced by the self-interaction term, which here is the pendular motion performed by $\mathbf{D}_1$.

\begin{figure*}
    \includegraphics[width=0.47\textwidth]{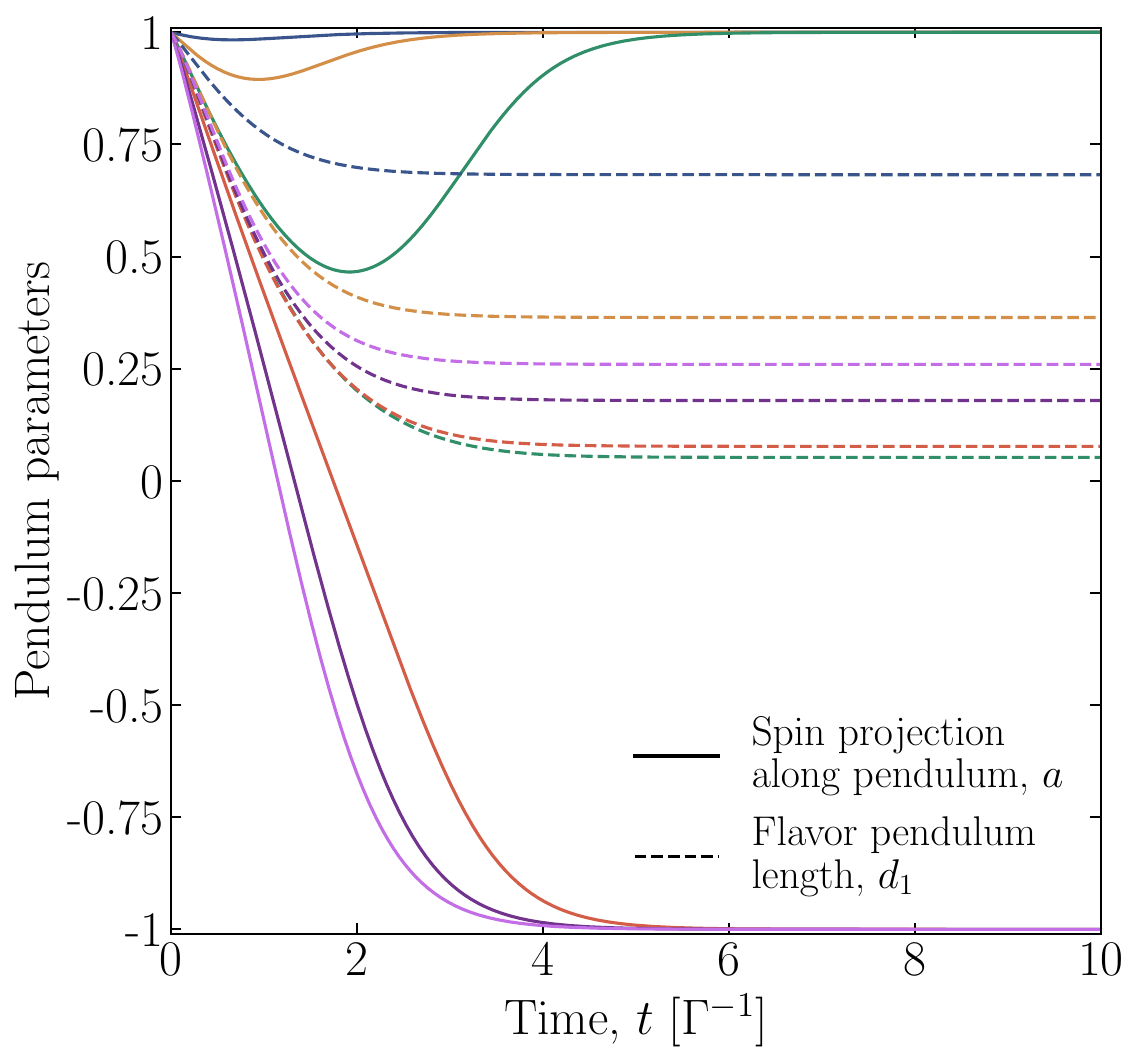}
    \kern1em
    \includegraphics[width=0.45\textwidth]{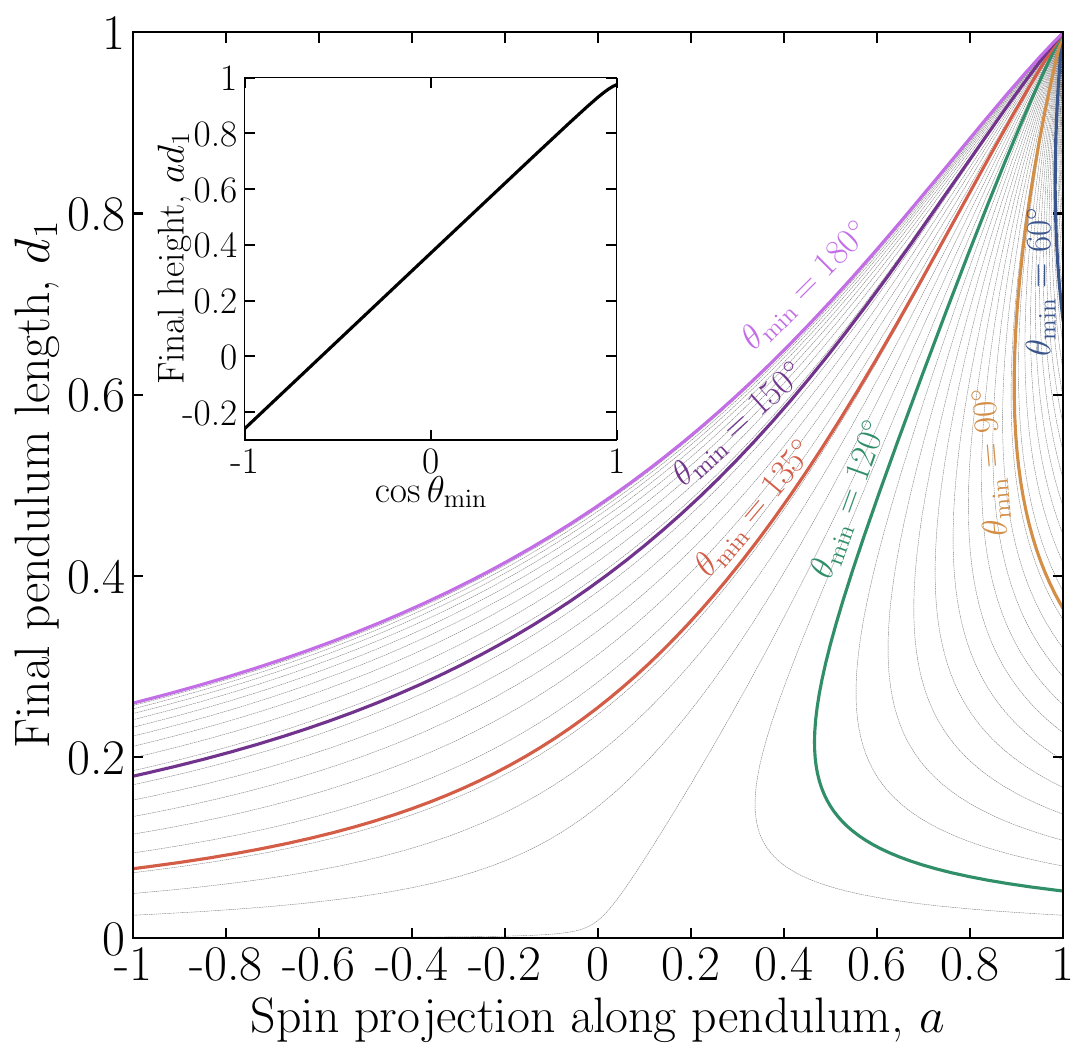}
    \caption{Evolution of the pendulum parameters obtained from the SDEs. \textit{Left:} Dynamical evolution of the pendulum spin along the axis $a=\bJ\cdot\bD_1/D_1 J_z$ and the dimensionless pendulum length $d_1=D_1/D_1(0)$. Different colors correspond to different values of $\theta_\mathrm{min}$, as identified in the right panel. \textit{Right:} Evolution of the pendulum parameters in the $a-d_1$ plane for various values of $\theta_\mathrm{min}$; we show in gray a sampling of the paths for uniformly sampled values of $\theta_\mathrm{min}$. In the inset, we show the final value of the pendulum height, $a d_1$, as a function of $\cos\theta_\mathrm{min}$.}\label{fig:pendulum_evolution}
\end{figure*}

From the EOMs, we derive the following evolution equations
\begin{subequations}
    \begin{eqnarray}
        \dot{J}_z&=&0,
        \\[1ex]
        \dot{D}_1&=&-\Gamma D_1\bigl(1-\overline{X^2}\bigr),
        \\
        \frac{d(\alpha D_1)}{dt}&=&-2\Gamma D_1\bigl(\alpha -J_z \overline{X}\bigr),
        \\
        \frac{dE}{dt}&=&-\Gamma\bigl(2E-J_z^2-2pD_0D_1\overline{X}\bigr),
    \end{eqnarray}
\end{subequations}
where $X=\cos\theta=\mathbf{D}_1\cdot\mathbf{e}_z/D_1$ is the deviation of the pendulum from the flavor direction and the overline denotes the average over a pendulum period.

This set of equations can be simplified by introducing the quantity $\Phi=\alpha D_0 D_1 p-J_z E$, whose derivative is obtained from the previous set as
\begin{equation}
   \dot{\Phi}=-2\Gamma\Phi-\Gamma J_z^3. 
\end{equation}
Since $J_z$ is conserved, this tells us that $\Phi$ relaxes to its asymptotic value exponentially. However, if the pendulum is initially in its upright position, as we always assume, and $E=\frac{1}{2}J_z^2+p D_0 D_1$, and so $\Phi(0)=-\frac{1}{2}J_z^3$. Therefore, $\Phi$ is conserved throughout the whole motion and is an adiabatic invariant, which means that
\begin{equation}
    E=\frac{J_z^2}{2}+\frac{\alpha}{J_z}D_0 D_1 p
\end{equation}
at every instant. Finally we have only two variables, $\alpha$ and $D_1$, because $J_z$ is constant and $E$ is determined at each instant by the conserved $\Phi$. To simplify the equations, we express everything in dimensionless variables $a=\alpha/J_z$ and $d_1=D_1/D_1(0)$. Initially, both $a$ and $d_1$ are equal to 1.

We redefine time units such that $\Gamma=1$. Now $a$ and $d_1$ obey the equations
\begin{subequations}\label{eq:renormgroup}
    \begin{eqnarray}
        \dot{a}&=&-a(1+\overline{X^2})+2\overline{X},
        \\
        \dot{d}_1&=&-d_1(1-\overline{X^2}),
    \end{eqnarray}
\end{subequations}
From the energy conservation equations, we find
\begin{eqnarray}
   \kern-2em &&\dot{X}^2={}
    \nonumber \\
    \kern-2em &&J_z^2\left\{\left[1-a^2+\frac{d_1(a-X)}{1+\cos\theta_\mathrm{min}}\right](1-X^2)-(1-aX)^2\right\}.
   \nonumber \\
    \kern-2em &&
\end{eqnarray}
Here $\cos\theta_\mathrm{min}$ is the maximum depth of flavor conversion, namely the cosine of the minimum zenith angle reached by the pendulum in its unperturbed state, defined as
\begin{equation}
    \cos\theta_\mathrm{min}=-1+\frac{J_z}{2pD_0 D_1(0)}.
\end{equation}
Therefore, we define
\begin{equation}
    I_n=\int\frac{X^n dX}{\sqrt{\left[1-a^2+\frac{d_1(a-X)}{1+\cos\theta_\mathrm{min}}\right](1-X^2)-(1-aX)^2}},
\end{equation}
where the integral is taken only over the region where the square root argument is positive and $-1<X<1$; one can check that this implies the integration range
\begin{equation}
    \frac{1-\sqrt{1+4d_1^2\xi^2-4ad_1\xi}}{2d_1\xi}<X<a,
\end{equation}
where $\xi=(1+\cos\theta_\mathrm{min})^{-1}$. With this definition, we write
\begin{equation}
    \overline{X^n}=\frac{I_n}{I_0}.
\end{equation}
This expresses the averages in Eqs.~\eqref{eq:renormgroup} explicitly in terms of $a$ and $d_1$. The dynamics of $a$ and $d_1$ is now expressed in the form of a closed set of equations, where at every next instant the new pendulum parameters depend only on the pendulum parameters at the previous instant, and on the control parameter $\cos\theta_\mathrm{min}$. 

Once more, the SDEs cannot be solved analytically, but offer considerable insight about the solutions. First, no terms of order $\mu$ appear in the equations. Second, only two quantities remain, namely the length of $\bD_1$ and the projection of the angular momentum over the pendulum length $\alpha$, which completely characterize the motion. Most of the original parameters of the problem have disappeared, including the collision rate $\Gamma$. This confirms the empirical findings of Ref.~\cite{Padilla-Gay:2022wck} that, provided $\Gamma\ll\mu$, the precise value of $\Gamma$ only determines the evolutionary speed towards a stationary state, but not the final state itself. Similarly, the SDEs depend only on the single parameter $\cos\theta_\mathrm{min}$. This proves the empirical finding that the final state only depends on $\cos\theta_\mathrm{min}$.

The SDEs determine directly the evolution of the pendulum parameters, without need to solve the EOMs for all polarization vectors. The evolution from integrating the SDEs is shown in Fig.~\ref{fig:pendulum_evolution}. The dynamics for a few selected values of $\cos\theta_\mathrm{min}$ is shown in the left panel. Initially, all curves start from $a(t=0)=1$ (pendulum in the upright position) and $d_1(t=0)=1$. For $\theta_\mathrm{min}$ very small, the pendulum is always damped to a vertical position with reduced length; in the final state, $a\to 1$, implying that the pendulum ends up vertical. As $\theta_\mathrm{min}$ increases, the final length $d_1$ eventually reaches~0. As $\theta_\mathrm{min}$ increases further, the final outcome is to invert the direction, which is vertical but points downward compared to the original direction. This is signaled by $a \to -1$ in the final state. 

The entire evolution can be compactly represented in the $a$--$d_1$ plane, as in the right panel of Fig.~\ref{fig:pendulum_evolution}. All curves start from the upper right corner ($a=1$, $d_1=1$), and their subsequent path depends on $\theta_\mathrm{min}$. The final length $d_1(t\to\infty)$ can be directly obtained, for each value of $\theta_\mathrm{min}$, by identifying the intersection of each path with the axis $a=1$ or $a=-1$, corresponding to the pendulum finally damped to the vertical position, either upwards or downwards respectively. By integrating the SDEs, we recover the linear relation between the final height of the pendulum and $\cos\theta_\mathrm{min}$ that was found in Ref.~\cite{Padilla-Gay:2022wck}, as we show in the inset of Fig.~\ref{fig:pendulum_evolution}.

At each instant, the pendulum parameters can be used to recover the angular distribution of the (anti)neutrino density matrices. The practical procedure is outlined in Appendix D of Ref.~\cite{Fiorillo:2023hlk}. Because the damped pendulum is always pushed toward its vertical position during flavor evolution, the off-diagonal terms of the final-state density matrices always vanish, as expected.
Nevertheless, the specific shape of the ELN angular distribution is in most cases non-trivial and specific scenarios need to be systematically explored in a full multi-angle framework; particular examples with ELN angular crossings are shown in Fig.2 of Ref.~\cite{Padilla-Gay:2022wck}. 

\section{Conclusions}\label{sec:conclusions}

We have presented a novel, approximate approach to solving the collective evolution of a homogeneous neutrino system that is subject to collisions with a background medium. The key element is the separation of the fast dynamics implied by the large scale $\mu$ of neutrino-neutrino refraction and the slow dynamics caused by the collision rate $\Gamma$. The justification for the separation of scales is the weak-damping limit $\Gamma\ll\mu$ that applies in all practical cases and means that between collisions, the system performs many oscillations, allowing one to use oscillation-averaged density matrices (or polarization vectors) to follow the slow evolution caused by collisions. One traditional example is the production of sterile neutrinos in the early universe or in SN cores, where the fast dynamics can be analytically integrated out. We have presented this case as a first warm-up exercise.

The novelty of our work consists in applying this method to much more complicated systems, where integrating out the fast dynamics leads to new slow-dynamics equations that we call SDEs, which themselves are nonlinear equations that require numerical solution, yet provide insights about the evolution and final state that are otherwise obscured by the fast dynamics. 

The approach of separating fast from slow dynamics in a controlled way by averaging over the cycle of a fast periodic motion, in the mechanical domain similar to Kapitza's pendulum~\cite{kapitza1965dynamical}, differs from other attempts to remove unnecessary detail from the EOMs. In statistical mechanics, one removes the motion of the microscopic degrees of freedom and studies macroscopic quantities such as temperature or pressure. Coarse graining over the microscopic degrees of freedom of a fast flavor system may lead to analogous  simplifications~\cite{Johns:2023jjt}, but the underlying assumption of the system relaxing to thermal equilibrium remains for the moment a conjecture.

We have studied two nontrivial applications, the collisional flavor instability (CFI) and the fast flavor pendulum with an energy-independent collision term. There are three main advantages compared to a full solution: (i)~A much coarser numerical step size. (ii)~A smaller number of variables in the slow dynamics; this shows up most clearly in the fast flavor pendulum, where the large set of polarization vectors reduce to two coupled differential equations for the pendulum length and spin. (iii)~Most importantly, the SDEs reveal certain aspects of the final state without actually solving them.

Notably in the case of CFIs, this approach allowed us to deduce the main features of the final distributions, namely equilibrium between the $e$ and $x$ flavor at low energies, and no flavor conversion at high energies. For the fast flavor pendulum, we proved that the final flavor composition depends only on the depth of the unperturbed pendulum, and not on $\Gamma$ or any other property, a conclusion that had been previously reached empirically.

For the CFI, there is a subtle point about the scale of fast dynamics. Usually, the scale of the neutrino-neutrino interaction energy is estimated as $\mu=\sqrt{2}G_{\rm F}n_\nu$. On the other hand, the relevant scale for the CFI derives from the length of the initial overall polarization vector and thus initially, $\mu_{\rm CFI}=\sqrt{2}G_{\rm F}\,(n_{\nu_e}-n_{\overline{\nu}_e}-n_{\overline{\nu}_x}+n_{\nu_x})$ is the relevant scale. Therefore, our method applies only when $\Gamma\ll\mu_{\rm CFI}$, whereas the opposite case $\mu_{\rm CFI}\ll\Gamma$ is sometimes termed the resonance-like regime~\cite{Xiong:2022zqz} in which a complete flavor swap may occur \cite{Kato:2023cig}. Conceivably, even in this case, our method could be applied in a later phase when the $z$ component of the global polarization vector has acquired a significant length.

In practice, our approach requires one to understand exactly the fast dynamics in the absence of collisions, so that the slow dynamics only touches those quantities that would be left invariant by the fast dynamics. This is why we had to restrict ourselves to homogeneous, azimuthally symmetric setups, for which the dynamics without collisions can be expressed analytically. However, if one were able to develop an understanding of the fast dynamics in inhomogeneous systems, our method could be applied to this more practically interesting case as well.

The impact of collisional flavor instabilities cannot be gauged by the limited information provided by linear stability analysis. A self-consistent solution should carefully incorporate matter-neutrino collisions and a multi-dimensional treatment of neutrino advection. Given the present difficulties in estimating flavor evolution in such complex environments, one possible way forward could be the approach presented in this work, where slowly-changing quantities can leverage some of the technical difficulties in the nonlinear regime. This new framework could pave the road toward understanding CFI in more complex systems that go beyond the common assumptions of isotropy and homogeneity.

\section*{Acknowledgements}

We are grateful to Julien Froustey for useful conversations and for pointing out a typo in our manuscript. DFGF is supported by the Villum Fonden under Project No.\ 29388 and the European Union's Horizon 2020 Research and Innovation Program under the Marie Sk{\l}odowska-Curie Grant Agreement No.\ 847523 ``\hbox{INTERACTIONS}.'' IPG acknowledges support from the U.S. Department of Energy under contract number DE-AC02-76SF00515. GR acknowledges support by the German Research Foundation (DFG) through the Collaborative Research Centre ``Neutrinos and Dark Matter in Astro and Particle Physics (NDM),'' Grant SFB-1258, and under Germany's Excellence Strategy through the Cluster of Excellence ORIGINS EXC-2094-390783311. 

\bibliographystyle{bibi}
\bibliography{Biblio}

\providecommand{\href}[2]{#2}\begingroup\raggedright\begin{thebibliography}{10}

\bibitem{Pantaleone:1992eq}
J.~T. Pantaleone, \emph{{Neutrino oscillations at high densities}},
  \href{https://doi.org/10.1016/0370-2693(92)91887-F}{\emph{Phys. Lett. B}
  {\bfseries 287} (1992) 128}.

\bibitem{Duan:2010bg}
H.~Duan, G.~M. Fuller and Y.-Z. Qian, \emph{{Collective Neutrino
  Oscillations}},
  \href{https://doi.org/10.1146/annurev.nucl.012809.104524}{\emph{Ann. Rev.
  Nucl. Part. Sci.} {\bfseries 60} (2010) 569}
  [\href{https://arxiv.org/abs/1001.2799}{{\ttfamily 1001.2799}}].

\bibitem{Mirizzi:2015eza}
A.~Mirizzi, I.~Tamborra, H.-T. Janka, N.~Saviano, K.~Scholberg, R.~Bollig,
  L.~H{\"u}depohl and S.~Chakraborty, \emph{{Supernova Neutrinos: Production,
  Oscillations and Detection}},
  \href{https://doi.org/10.1393/ncr/i2016-10120-8}{\emph{Riv. Nuovo Cim.}
  {\bfseries 39} (2016) 1} [\href{https://arxiv.org/abs/1508.00785}{{\ttfamily
  1508.00785}}].

\bibitem{Tamborra:2020cul}
I.~Tamborra and S.~Shalgar, \emph{{New Developments in Flavor Evolution of a
  Dense Neutrino Gas}},
  \href{https://doi.org/10.1146/annurev-nucl-102920-050505}{\emph{Ann. Rev.
  Nucl. Part. Sci.} {\bfseries 71} (2021) 165}
  [\href{https://arxiv.org/abs/2011.01948}{{\ttfamily 2011.01948}}].

\bibitem{Capozzi:2022slf}
F.~Capozzi and N.~Saviano, \emph{{Neutrino Flavor Conversions in High-Density
  Astrophysical and Cosmological Environments}},
  \href{https://doi.org/10.3390/universe8020094}{\emph{Universe} {\bfseries 8}
  (2022) 94} [\href{https://arxiv.org/abs/2202.02494}{{\ttfamily 2202.02494}}].

\bibitem{Richers:2022zug}
S.~Richers and M.~Sen, \emph{{Fast Flavor Transformations}},
  \href{https://doi.org/10.1007/978-981-15-8818-1_125-1}{\emph{Handbook of
  Nuclear Physics} (2022) 1}
  [\href{https://arxiv.org/abs/2207.03561}{{\ttfamily 2207.03561}}].

\bibitem{Harris:1980zi}
R.~A. Harris and L.~Stodolsky, \emph{{Two State Systems in Media and ``Turing's
  Paradox''}}, \href{https://doi.org/10.1016/0370-2693(82)90169-1}{\emph{Phys.
  Lett. B} {\bfseries 116} (1982) 464}.

\bibitem{Johns:2021qby}
L.~Johns, \emph{{Collisional Flavor Instabilities of Supernova Neutrinos}},
  \href{https://doi.org/10.1103/PhysRevLett.130.191001}{\emph{Phys. Rev. Lett.}
  {\bfseries 130} (2023) 191001}
  [\href{https://arxiv.org/abs/2104.11369}{{\ttfamily 2104.11369}}].

\bibitem{Xiong:2022zqz}
Z.~Xiong, L.~Johns, M.-R. Wu and H.~Duan, \emph{{Collisional flavor instability
  in dense neutrino gases}},
  \href{https://doi.org/10.1103/PhysRevD.108.083002}{\emph{Phys. Rev. D}
  {\bfseries 108} (2023) 083002}
  [\href{https://arxiv.org/abs/2212.03750}{{\ttfamily 2212.03750}}].

\bibitem{Fiorillo:2024}
D.~F.~G. Fiorillo, I.~Padilla-Gay and G.~G. Raffelt, \emph{{Collisional Flavor
  Instability for Pedestrians}},  2024.
\newblock Work in Progress.

\bibitem{Liu:2023pjw}
J.~Liu, M.~Zaizen and S.~Yamada, \emph{{Systematic study of the resonancelike
  structure in the collisional flavor instability of neutrinos}},
  \href{https://doi.org/10.1103/PhysRevD.107.123011}{\emph{Phys. Rev. D}
  {\bfseries 107} (2023) 123011}
  [\href{https://arxiv.org/abs/2302.06263}{{\ttfamily 2302.06263}}].

\bibitem{Lin:2022dek}
Y.-C. Lin and H.~Duan, \emph{{Collision-induced flavor instability in dense
  neutrino gases with energy-dependent scattering}},
  \href{https://doi.org/10.1103/PhysRevD.107.083034}{\emph{Phys. Rev. D}
  {\bfseries 107} (2023) 083034}
  [\href{https://arxiv.org/abs/2210.09218}{{\ttfamily 2210.09218}}].

\bibitem{Johns:2022yqy}
L.~Johns and Z.~Xiong, \emph{{Collisional instabilities of neutrinos and their
  interplay with fast flavor conversion in compact objects}},
  \href{https://doi.org/10.1103/PhysRevD.106.103029}{\emph{Phys. Rev. D}
  {\bfseries 106} (2022) 103029}
  [\href{https://arxiv.org/abs/2208.11059}{{\ttfamily 2208.11059}}].

\bibitem{Padilla-Gay:2022wck}
I.~Padilla-Gay, I.~Tamborra and G.~G. Raffelt, \emph{{Neutrino fast flavor
  pendulum. II. Collisional damping}},
  \href{https://doi.org/10.1103/PhysRevD.106.103031}{\emph{Phys. Rev. D}
  {\bfseries 106} (2022) 103031}
  [\href{https://arxiv.org/abs/2209.11235}{{\ttfamily 2209.11235}}].

\bibitem{Wolfenstein:1977ue}
L.~Wolfenstein, \emph{{Neutrino oscillations in matter}},
  \href{https://doi.org/10.1103/PhysRevD.17.2369}{\emph{Phys. Rev. D}
  {\bfseries 17} (1978) 2369}.

\bibitem{Wolfenstein:1979ni}
L.~Wolfenstein, \emph{{Neutrino Oscillations and Stellar Collapse}},
  \href{https://doi.org/10.1103/PhysRevD.20.2634}{\emph{Phys. Rev. D}
  {\bfseries 20} (1979) 2634}.

\bibitem{Mikheyev:1985zog}
S.~P. Mikheyev and A.~Y. Smirnov, \emph{{Resonance Amplification of
  Oscillations in Matter and Spectroscopy of Solar Neutrinos}}, {\emph{Sov. J.
  Nucl. Phys.} {\bfseries 42} (1985) 913}.

\bibitem{Padilla-Gay:2021haz}
I.~Padilla-Gay, I.~Tamborra and G.~G. Raffelt, \emph{{Neutrino Flavor Pendulum
  Reloaded: The Case of Fast Pairwise Conversion}},
  \href{https://doi.org/10.1103/PhysRevLett.128.121102}{\emph{Phys. Rev. Lett.}
  {\bfseries 128} (2022) 121102}
  [\href{https://arxiv.org/abs/2109.14627}{{\ttfamily 2109.14627}}].

\bibitem{Fiorillo:2023mze}
D.~F.~G. Fiorillo and G.~G. Raffelt, \emph{{Slow and fast collective neutrino
  oscillations: Invariants and reciprocity}},
  \href{https://doi.org/10.1103/PhysRevD.107.043024}{\emph{Phys. Rev. D}
  {\bfseries 107} (2023) 043024}
  [\href{https://arxiv.org/abs/2301.09650}{{\ttfamily 2301.09650}}].

\bibitem{Manohar:1986gj}
A.~Manohar, \emph{{Statistical Mechanics of Noninteracting Particles}},
  \href{https://doi.org/10.1016/0370-2693(87)90310-8}{\emph{Phys. Lett. B}
  {\bfseries 186} (1987) 370}.

\bibitem{Barbieri:1989ti}
R.~Barbieri and A.~Dolgov, \emph{{Bounds on Sterile-neutrinos from
  Nucleosynthesis}},
  \href{https://doi.org/10.1016/0370-2693(90)91203-N}{\emph{Phys. Lett. B}
  {\bfseries 237} (1990) 440}.

\bibitem{Cline:1991zb}
J.~M. Cline, \emph{{Constraints on almost Dirac neutrinos from
  neutrino-antineutrino oscillations}},
  \href{https://doi.org/10.1103/PhysRevLett.68.3137}{\emph{Phys. Rev. Lett.}
  {\bfseries 68} (1992) 3137}.

\bibitem{Enqvist:1991qj}
K.~Enqvist, K.~Kainulainen and M.~J. Thomson, \emph{{Stringent cosmological
  bounds on inert neutrino mixing}},
  \href{https://doi.org/10.1016/0550-3213(92)90442-E}{\emph{Nucl. Phys. B}
  {\bfseries 373} (1992) 498}.

\bibitem{Dodelson:1993je}
S.~Dodelson and L.~M. Widrow, \emph{{Sterile-neutrinos as dark matter}},
  \href{https://doi.org/10.1103/PhysRevLett.72.17}{\emph{Phys. Rev. Lett.}
  {\bfseries 72} (1994) 17}
  [\href{https://arxiv.org/abs/hep-ph/9303287}{{\ttfamily hep-ph/9303287}}].

\bibitem{Abazajian:2001nj}
K.~Abazajian, G.~M. Fuller and M.~Patel, \emph{{Sterile neutrino hot, warm, and
  cold dark matter}},
  \href{https://doi.org/10.1103/PhysRevD.64.023501}{\emph{Phys. Rev. D}
  {\bfseries 64} (2001) 023501}
  [\href{https://arxiv.org/abs/astro-ph/0101524}{{\ttfamily
  astro-ph/0101524}}].

\bibitem{Hannestad:2012ky}
S.~Hannestad, I.~Tamborra and T.~Tram, \emph{{Thermalisation of light sterile
  neutrinos in the early universe}},
  \href{https://doi.org/10.1088/1475-7516/2012/07/025}{\emph{JCAP} {\bfseries
  07} (2012) 025} [\href{https://arxiv.org/abs/1204.5861}{{\ttfamily
  1204.5861}}].

\bibitem{Abazajian:2012ys}
K.~N. Abazajian et~al., \emph{{Light Sterile Neutrinos: A White Paper}},
  \href{https://arxiv.org/abs/1204.5379}{{\ttfamily 1204.5379}}.

\bibitem{Kainulainen:1990bn}
K.~Kainulainen, J.~Maalampi and J.~T. Peltoniemi, \emph{{Inert neutrinos in
  supernovae}}, \href{https://doi.org/10.1016/0550-3213(91)90354-Z}{\emph{Nucl.
  Phys. B} {\bfseries 358} (1991) 435}.

\bibitem{Raffelt:1992bs}
G.~Raffelt and G.~Sigl, \emph{{Neutrino flavor conversion in a supernova
  core}}, \href{https://doi.org/10.1016/0927-6505(93)90020-E}{\emph{Astropart.
  Phys.} {\bfseries 1} (1993) 165}
  [\href{https://arxiv.org/abs/astro-ph/9209005}{{\ttfamily
  astro-ph/9209005}}].

\bibitem{Shi:1993ee}
X.~Shi and G.~Sigl, \emph{{A Type II supernovae constraint on $\nu_e$--$\nu_s$
  mixing}}, \href{https://doi.org/10.1016/0370-2693(94)91232-7}{\emph{Phys.
  Lett. B} {\bfseries 323} (1994) 360}
  [\href{https://arxiv.org/abs/hep-ph/9312247}{{\ttfamily hep-ph/9312247}}].
  Erratum: \href{https://doi.org/10.1016/0370-2693(94)90233-X} {{\em Phys.
  Lett. B} {\bf 324}, 516 (1994)}.

\bibitem{Nunokawa:1997ct}
H.~Nunokawa, J.~T. Peltoniemi, A.~Rossi and J.~W.~F. Valle, \emph{{Supernova
  bounds on resonant active sterile neutrino conversions}},
  \href{https://doi.org/10.1103/PhysRevD.56.1704}{\emph{Phys. Rev. D}
  {\bfseries 56} (1997) 1704}
  [\href{https://arxiv.org/abs/hep-ph/9702372}{{\ttfamily hep-ph/9702372}}].

\bibitem{Hidaka:2007se}
J.~Hidaka and G.~M. Fuller, \emph{{Sterile Neutrino-Enhanced Supernova
  Explosions}}, \href{https://doi.org/10.1103/PhysRevD.76.083516}{\emph{Phys.
  Rev. D} {\bfseries 76} (2007) 083516}
  [\href{https://arxiv.org/abs/0706.3886}{{\ttfamily 0706.3886}}].

\bibitem{Raffelt:2011nc}
G.~G. Raffelt and S.~Zhou, \emph{{Supernova bound on keV-mass sterile neutrinos
  reexamined}}, \href{https://doi.org/10.1103/PhysRevD.83.093014}{\emph{Phys.
  Rev. D} {\bfseries 83} (2011) 093014}
  [\href{https://arxiv.org/abs/1102.5124}{{\ttfamily 1102.5124}}].

\bibitem{Arguelles:2016uwb}
C.~A. Arg\"uelles, V.~Brdar and J.~Kopp, \emph{{Production of keV Sterile
  Neutrinos in Supernovae: New Constraints and Gamma Ray Observables}},
  \href{https://doi.org/10.1103/PhysRevD.99.043012}{\emph{Phys. Rev. D}
  {\bfseries 99} (2019) 043012}
  [\href{https://arxiv.org/abs/1605.00654}{{\ttfamily 1605.00654}}].

\bibitem{Suliga:2019bsq}
A.~M. Suliga, I.~Tamborra and M.-R. Wu, \emph{{Tau lepton asymmetry by sterile
  neutrino emission -- Moving beyond one-zone supernova models}},
  \href{https://doi.org/10.1088/1475-7516/2019/12/019}{\emph{JCAP} {\bfseries
  12} (2019) 019} [\href{https://arxiv.org/abs/1908.11382}{{\ttfamily
  1908.11382}}].

\bibitem{Suliga:2020vpz}
A.~M. Suliga, I.~Tamborra and M.-R. Wu, \emph{{Lifting the core-collapse
  supernova bounds on keV-mass sterile neutrinos}},
  \href{https://doi.org/10.1088/1475-7516/2020/08/018}{\emph{JCAP} {\bfseries
  08} (2020) 018} [\href{https://arxiv.org/abs/2004.11389}{{\ttfamily
  2004.11389}}].

\bibitem{Froustey:2020mcq}
J.~Froustey, C.~Pitrou and M.~C. Volpe, \emph{{Neutrino decoupling including
  flavour oscillations and primordial nucleosynthesis}},
  \href{https://doi.org/10.1088/1475-7516/2020/12/015}{\emph{JCAP} {\bfseries
  12} (2020) 015} [\href{https://arxiv.org/abs/2008.01074}{{\ttfamily
  2008.01074}}].

\bibitem{Froustey:2021azz}
J.~Froustey and C.~Pitrou, \emph{{Primordial neutrino asymmetry evolution with
  full mean-field effects and collisions}},
  \href{https://doi.org/10.1088/1475-7516/2022/03/065}{\emph{JCAP} {\bfseries
  03} (2022) 065} [\href{https://arxiv.org/abs/2110.11889}{{\ttfamily
  2110.11889}}].

\bibitem{Dolgov:1980cq}
A.~D. Dolgov, \emph{{Neutrinos in the early universe}}, {\emph{Sov. J. Nucl.
  Phys.} {\bfseries 33} (1981) 700}. [{\em Yad.\ Fiz.} {\bf 33} (1981) 1309].

\bibitem{Sigl:1993ctk}
G.~Sigl and G.~Raffelt, \emph{{General kinetic description of relativistic
  mixed neutrinos}},
  \href{https://doi.org/10.1016/0550-3213(93)90175-O}{\emph{Nucl. Phys. B}
  {\bfseries 406} (1993) 423}.

\bibitem{Kuramoto1975}
Y.~Kuramoto, \emph{Self-entrainment of a population of coupled non-linear
  oscillators},  in \emph{International Symposium on Mathematical Problems in
  Theoretical Physics} (H.~Araki, ed.), (Berlin, Heidelberg), pp.~420--422,
  Springer Berlin Heidelberg, 1975.

\bibitem{Kuramoto2003}
Y.~Kuramoto, \emph{Chemical oscillations, waves, and turbulence}, Chemistry
  Series. Dover Publications, 2003.

\bibitem{Pantaleone:1998xi}
J.~T. Pantaleone, \emph{{Stability of incoherence in an isotropic gas of
  oscillating neutrinos}},
  \href{https://doi.org/10.1103/PhysRevD.58.073002}{\emph{Phys. Rev. D}
  {\bfseries 58} (1998) 073002}.

\bibitem{Johns:2019izj}
L.~Johns, H.~Nagakura, G.~M. Fuller and A.~Burrows, \emph{{Neutrino
  oscillations in supernovae: angular moments and fast instabilities}},
  \href{https://doi.org/10.1103/PhysRevD.101.043009}{\emph{Phys. Rev. D}
  {\bfseries 101} (2020) 043009}
  [\href{https://arxiv.org/abs/1910.05682}{{\ttfamily 1910.05682}}].

\bibitem{Johns:2020qsk}
L.~Johns, H.~Nagakura, G.~M. Fuller and A.~Burrows, \emph{{Fast oscillations,
  collisionless relaxation, and spurious evolution of supernova neutrino
  flavor}}, \href{https://doi.org/10.1103/PhysRevD.102.103017}{\emph{Phys. Rev.
  D} {\bfseries 102} (2020) 103017}
  [\href{https://arxiv.org/abs/2009.09024}{{\ttfamily 2009.09024}}].

\bibitem{Fiorillo:2023hlk}
D.~F.~G. Fiorillo and G.~G. Raffelt, \emph{{Flavor solitons in dense neutrino
  gases}}, \href{https://doi.org/10.1103/PhysRevD.107.123024}{\emph{Phys. Rev.
  D} {\bfseries 107} (2023) 123024}
  [\href{https://arxiv.org/abs/2303.12143}{{\ttfamily 2303.12143}}].

\bibitem{Johns:2022bmu}
L.~Johns and H.~Nagakura, \emph{{Self-consistency in models of neutrino
  scattering and fast flavor conversion}},
  \href{https://doi.org/10.1103/PhysRevD.106.043031}{\emph{Phys. Rev. D}
  {\bfseries 106} (2022) 043031}
  [\href{https://arxiv.org/abs/2206.09225}{{\ttfamily 2206.09225}}].

\bibitem{kapitza1965dynamical}
P.~L. Kapitza, \emph{Dynamical stability of a pendulum when its point of
  suspension vibrates, and pendulum with a vibrating suspension},
  {\emph{Collected papers of PL Kapitza} {\bfseries 2} (1965) 714}.

\bibitem{Johns:2023jjt}
L.~Johns, \emph{{Thermodynamics of oscillating neutrinos}},
  \href{https://arxiv.org/abs/2306.14982}{{\ttfamily 2306.14982}}.

\bibitem{Kato:2023cig}
C.~Kato, H.~Nagakura and L.~Johns, \emph{{Collisional flavor swap with neutrino
  self-interactions}},  \href{https://arxiv.org/abs/2309.02619}{{\ttfamily
  2309.02619}}.

\end{thebibliography}\endgroup

\end{document}